# Fractional Electromagnetic Field Theory and Its Applications

Hosein Nasrolahpour[*]
School of Computational Science and Engineering, University of Tehran, Tehran, Iran.

## Abstract

Fractional electromagnetic field theory describes electromagnetic wave propagation through the complex, nonlocal, dissipative, fractal and also recent artificially engineered materials know as fractional metamaterials. In this theory using the fractional Maxwell's equations we are also able to consider the well-known effect of electromagnetic memory. In this review we present some applications of the powerful theory of fractional electrodynamics.

**Keywords:** Fractional Calculus; Fractional Electrodynamics; Fractional Maxwell's Equations; Fractional Metamaterials; Fractional Drude Model; Electromagnetic Memory Effect; Fotonics

## 1. Introduction

Fractional electrodynamics and the theory of fractional electromagnetics are based on the fractional Maxwell's equations and describe electromagnetic wave propagation through the complex, nonlocal, dissipative, fractal and also recent artificially engineered materials known as metamaterials [1]. In this theory we are also able to consider the well-known effect of electromagnetic memory. In this review we first present a short introduction to the fractional vector calculus as mathematical preliminary for the fractional electrodynamics. In the next section we start to discuss about the fractional electrodynamics by introducing the so called fractional maxwell's equations. In this section we investigate the time evolution of the fractional electromagnetic waves by using the time fractional Maxwell's equations. In particular, we show that electromagnetic plane wave has an amplitude which exhibits an algebraic decay , at asymptotically large times. In section 4 we consider our recently proposed notion of fractional electromagnetic metamaterials [2]. For this purpose, we first introduce the new concept of fractional order metamaterials then we derive governing electromagnetic equations to model this new type of electromagnetic metamaterials. In the next section we discuss about the fractional dynamics of optical and photonic phenomena in complex media. We called this new field of research as fotonics which can be considered as a generalization of the conventional optics and photonics [3]. As an example of its application, we investigate the physics of random lasers in this framework.

## 2. Fractional Vector Calculus

Fractional calculus (FC) is very useful and powerful tool for modeling phenomena which exhibit space time nonlocality [4]. In this section we present fractional generalization of vector calculus which are needed for building our new model [5-7 and refs. therein]. For this aim, we use the well-known Caputo partial fractional derivatives for the time derivatives and the Riesz fractional derivatives for the space ones and defining the fractional Laplacian. Left (forward) and right (backward) Caputo (RL) partial fractional derivatives of order $\alpha_\mu$, $\beta_\mu$ (which are positive real or even complex numbers) of a real valued function $f$ of $d+1$ real variables $x^0, x^1, ..., x^d$ with respect to $x_\mu$ are as follow:

$$_{a_\mu}^C\partial_\mu^{\alpha_\mu} f(x^0,...,x^d) = \frac{1}{\Gamma(n_\mu - \alpha_\mu)} \int_{a_\mu}^x \frac{\partial_{x^\mu}^{n_\mu} f(x^0,...,x^{\mu-1}, u, x^{\mu+1},...,x^d)}{(x^\mu - u)^{1+\alpha_\mu - n_\mu}} du \quad \text{(left Caputo)} \quad (1)$$

$$_\mu^C\partial_{b_\mu}^{\beta_\mu} f(x^0,...,x^d) = \frac{(-1)^{n_\mu}}{\Gamma(n_\mu - \beta_\mu)} \int_x^{b_\mu} \frac{\partial_{x^\mu}^{n_\mu} f(x^0,...,x^{\mu-1}, u, x^{\mu+1},...,x^d)}{(u - x^\mu)^{1+\beta_\mu - n_\mu}} du \quad \text{(right Caputo)} \quad (2)$$

where $\partial_{x^\mu}^{n_\mu}$ is the ordinary partial derivative of integer order $n$ with respect to the variable $x$ and $a_\mu$, $b_\mu$ are real number which define the domain [4 and refs. therein]. Also, for the Riesz fractional derivative of order $\alpha_\mu$ we have:

$$^R\partial^{\alpha_\mu} f(x^0,...,x^d) = -\frac{1}{2\cos(\pi\alpha_\mu/2)\Gamma(n_\mu - \alpha_\mu)} \partial_{x^\mu}^{n_\mu} (\int_{-\infty}^{x^\mu} \frac{f(x^0,...,x^{\mu-1},u,x^{\mu+1},...,x^d)}{(x^\mu - u)^{1+\alpha_\mu - n_\mu}} du + \int_{x^\mu}^{\infty} \frac{f(x^0,...,x^{\mu-1},u,x^{\mu+1},...,x^d)}{(u - x^\mu)^{1+\alpha_\mu - n_\mu}} du)$$

(3)

[*]Correspondence: Hosein Nasrolahpour, University of Tehran, Tehran, Iran.
E-Mail: hnasrolahpour@gmail.com & hnasrolahpour@ut.ac.ir

Using this definition, we can easily introduce fractional curl operator as:
$$(\vec{\nabla}\times)^\alpha \vec{F} = \hat{e}_\rho \varepsilon_{\rho\mu\nu} \partial_\mu^\alpha F_\nu \quad (4)$$

We can also derive a relation for the double curl operation in the following form:
$$(\vec{\nabla}\times)^\alpha ((\vec{\nabla}\times)^\alpha \vec{F}) = \hat{e}_\rho \varepsilon_{\rho\mu\nu} \varepsilon_{\nu\kappa\sigma} \partial_\mu^\alpha \partial_\kappa^\alpha F_\sigma \quad (5)$$

## 3. Fractional electrodynamics

In classical electrodynamics, behavior of electric fields ($\vec{E}$), magnetic fields ($\vec{B}$) and their relations to their sources, charge density ($\rho(\vec{r},t)$) and current density ($\vec{j}(\vec{r},t)$), are described by the following Maxwell's equations:

$$\vec{\nabla}.\vec{E} = \frac{4\pi}{\varepsilon} \rho(\vec{r},t) \quad (6)$$

$$\vec{\nabla}.\vec{B} = 0 \quad (7)$$

$$\vec{\nabla}\times\vec{E} = -\frac{1}{c}\frac{\partial \vec{B}}{\partial t} \quad (8)$$

$$\vec{\nabla}\times\vec{B} = \frac{4\pi\mu}{c}\vec{j}(\vec{r},t) + \frac{\varepsilon\mu}{c}\frac{\partial \vec{E}}{\partial t} \quad (9)$$

Where $\varepsilon$ and $\mu$ are electric permittivity and magnetic permeability, respectively. Now, introducing the potentials, vector $\vec{A}(x_i,t)$ and scalar $\varphi(x_i,t)$

$$\vec{B} = \vec{\nabla} \times \vec{A} \tag{10}$$

$$\vec{E} = -\frac{1}{c}\frac{\partial \vec{A}}{\partial t} - \vec{\nabla}\varphi \tag{11}$$

and using the Lorenz gauge condition we obtain the following decoupled differential equations for the potentials:

$$\Delta \vec{A}(\vec{r},t) - \frac{\varepsilon\mu}{c^2}\frac{\partial^2 \vec{A}(\vec{r},t)}{\partial t^2} = -\frac{4\pi}{c}\vec{j}(\vec{r},t) \tag{12}$$

$$\Delta \varphi(\vec{r},t) - \frac{\varepsilon\mu}{c^2}\frac{\partial^2 \varphi(\vec{r},t)}{\partial t^2} = -\frac{4\pi}{\varepsilon}\rho(\vec{r},t) \tag{13}$$

where $\frac{\varepsilon\mu}{c^2} = \frac{1}{v^2}$ and $v$ is the velocity of the wave. Furthermore, for a particle motion with charge q in the presence of electric and magnetic field we can write the Lorentz force as

$$\vec{F}_L = q(\vec{E} + \vec{v} \times \vec{B}) \tag{14}$$

here, $v$ is the particle's velocity. In terms of scalar and vector potentials, Eq. (10, 11) we may write the Lorentz force as

$$\vec{F}_L = q(-\frac{1}{c}\frac{\partial \vec{A}}{\partial t} - \vec{\nabla}\varphi + \vec{v} \times (\vec{\nabla} \times \vec{A})) \tag{15}$$

As we can see in Refs. [8-10], in classical mechanics, the fractional formalism leads to relaxation and oscillation processes that exhibit memory and delay. This fractional nonlocal formalism is also applicable on materials and media that have electromagnetic memory properties. So, the generalized fractional Maxwell's equations can give us new models that can be used in these complex systems. Up to now, several different kinds of fractional electrodynamics based on the different approaches to fractional vector calculus have been investigated [11-16]. For instance, a fractional-dimensional space approach to the electrodynamics is presented [11] using the fractional-dimensional Laplacian operator $\nabla_D^2$ [17, 18]:

$$\nabla_D^2 = \frac{\partial^2}{\partial x^2} + \frac{\alpha_1 - 1}{x}\frac{\partial}{\partial x} + \frac{\partial^2}{\partial y^2} + \frac{\alpha_2 - 1}{y}\frac{\partial}{\partial y} + \frac{\partial^2}{\partial z^2} + \frac{\alpha_3 - 1}{z}\frac{\partial}{\partial z} \tag{16}$$

where, three parameters $(0 < \alpha_1 \leq 1,\ 0 < \alpha_2 \leq 1\ and\ 0 < \alpha_3 \leq 1)$ are used to describe the measure distribution of space where each one is acting independently on a single coordinate and the total dimension of the system is $D = \alpha_1 + \alpha_2 + \alpha_3$.

Also, in [19] spacetime fractional Maxwell's equations are presented using the fractional operators defined in Eq. (1,3) as:

$$\sigma_\alpha \vec{\nabla}^\alpha \cdot \vec{E} = \frac{4\pi}{\varepsilon}\rho(\vec{r},t) \tag{17}$$

$$\sigma_\alpha \vec{\nabla}^\alpha \cdot \vec{B} = 0 \tag{18}$$

$$\sigma_\alpha \vec{\nabla}^\alpha \times \vec{E} = -\frac{\eta_\alpha}{c}\frac{\partial^\alpha \vec{B}}{\partial t^\alpha} \tag{19}$$

$$\sigma_\alpha \vec{\nabla}^\alpha \times \vec{B} = \frac{4\pi\mu}{c} \vec{j}(\vec{r},t) + \frac{\eta_\alpha \varepsilon \mu}{c} \frac{\partial^\alpha \vec{E}}{\partial t^\alpha} \tag{20}$$

However, in this section we study a new approach in this area [20]. The idea is in fact, to write the ordinary differential wave equations in the fractional form with respect to $t$, by replacing the time derivative with a fractional derivative of order $\alpha$ $(0 < \alpha \leq 1)$ namely:

$$\vec{\nabla} \cdot \vec{E} = \frac{4\pi}{\varepsilon} \rho(\vec{r},t) \tag{21}$$

$$\vec{\nabla} \cdot \vec{B} = 0 \tag{22}$$

$$\vec{\nabla} \times \vec{E} = -\frac{1}{c} \frac{1}{\eta^{1-\alpha}} \frac{\partial^\alpha \vec{B}}{\partial t^\alpha} \tag{23}$$

$$\vec{\nabla} \times \vec{B} = \frac{4\pi\mu}{c} \vec{j}(\vec{r},t) + \frac{\varepsilon\mu}{c} \frac{1}{\eta^{1-\alpha}} \frac{\partial^\alpha \vec{E}}{\partial t^\alpha} \tag{24}$$

And the Eq. (10,11) become

$$\vec{B} = \vec{\nabla} \times \vec{A} \tag{25}$$

$$\vec{E} = -\frac{1}{c\eta^{1-\alpha}} \frac{\partial^\alpha \vec{A}}{\partial^\alpha t} - \vec{\nabla}\varphi \tag{26}$$

And the Lorentz force Eq. (15) becomes

$$\vec{F}_L = q\left(-\frac{1}{c\eta^{1-\alpha}} \frac{\partial^\alpha \vec{A}}{\partial^\alpha t} - \vec{\nabla}\varphi + \vec{v} \times (\vec{\nabla} \times \vec{A})\right) \tag{27}$$

A simple example of application of the Eq. (27) is provided in Appendix A.

Also, note that we have introduced an arbitrary quantity $\eta$ with dimension of [second] to ensure that all quantities have correct dimensions.

Now we can apply Lorentz gauge condition to obtain the corresponding time fractional wave equations for the potentials

$$\Delta \vec{A}(\vec{r},t) - \frac{\varepsilon\mu}{c^2} \frac{1}{\eta^{2(1-\alpha)}} \frac{\partial^{2\alpha} \vec{A}(\vec{r},t)}{\partial t^{2\alpha}} = -\frac{4\pi}{c} \vec{j}(\vec{r},t) \tag{28}$$

$$\Delta \varphi(\vec{r},t) - \frac{\varepsilon\mu}{c^2} \frac{1}{\eta^{2(1-\alpha)}} \frac{\partial^{2\alpha} \varphi(\vec{r},t)}{\partial t^{2\alpha}} = -\frac{4\pi}{\varepsilon} \rho(\vec{r},t) \tag{29}$$

If $\rho = 0$ and $\vec{j} = 0$, we have the homogeneous fractional differential equations

$$\Delta \vec{A}(\vec{r},t) - \frac{\varepsilon\mu}{c^2} \frac{1}{\eta^{2(1-\alpha)}} \frac{\partial^{2\alpha} \vec{A}(\vec{r},t)}{\partial t^{2\alpha}} = 0 \tag{30}$$

$$\Delta \varphi(\vec{r},t) - \frac{\varepsilon\mu}{c^2} \frac{1}{\eta^{2(1-\alpha)}} \frac{\partial^{2\alpha} \varphi(\vec{r},t)}{\partial t^{2\alpha}} = 0 \tag{31}$$

We are interested in the analysis of the electromagnetic fields starting from the equations. Now we can write the fractional equations in following compact form

$$\frac{\partial^2 Z(x,t)}{\partial x^2} - \frac{\varepsilon\mu}{c^2} \frac{1}{\eta^{2(1-\alpha)}} \frac{\partial^{2\alpha} Z(x,t)}{\partial t^{2\alpha}} = 0 \tag{32}$$

where $Z(x,t)$ represents both $\vec{A}(\vec{r},t)$ and $\varphi(\vec{r},t)$. We consider a polarized electromagnetic wave, then $A_x = 0, A_y \neq 0, A_z \neq 0$. A particular solution of this equation may be found in the form

$$Z(x,t) = Z_0 e^{-ikx} u(t) \tag{33}$$

where $k$ is the wave vector in the $x$ direction and $Z_0$ is a constant. Substituting Eq. (33) into Eq. (32) we obtain

$$\frac{d^{2\alpha} u(t)}{dt^{2\alpha}} + \Omega_f^2 u(t) = 0 \tag{34}$$

where

$$\Omega_f^2 = v^2 k^2 \eta^{2(1-\alpha)} = \Omega^2 \eta^{2(1-\alpha)} \tag{35}$$

and $\Omega$ is the fundamental frequency of the electromagnetic wave. Using the Laplace integral transformations:

$$L\{{}_0^c D_t^\alpha f(t)\} = s^\alpha F(s) - \sum_{m=0}^{n-1} s^{\alpha-m-1} f^{(m)}(0) \tag{36}$$

where, $F(s)$ is the Laplace transform of $f(t)$ one obtains the solutions:

$$u(t) = u(0) E_{2\alpha}(-\Omega_f^2 t^{2\alpha}) \tag{37}$$

for the case of $0 < \alpha < \frac{1}{2}$ and

$$u(t) = u(0) E_{2\alpha}(-\Omega_f^2 t^{2\alpha}) + t u'(0) E_{2\alpha,2}(-\Omega_f^2 t^{2\alpha}) \tag{38}$$

for the case of $\frac{1}{2} < \alpha < 1$. So with the boundary conditions

$$u(0) = 1 \text{ and } u'(0) = 0 \tag{39}$$

the general solution of the Eq. (34) may be

$$u(t) = E_{2\alpha}(-\Omega_f^2 t^{2\alpha}) \tag{40}$$

where

$$E_\alpha(z) = \sum_{k=0}^{\infty} \frac{z^k}{\Gamma(1+\alpha k)} \tag{41}$$

is one-parameter Mittag-Leffler function. Substituting the Eq. (40) in Eq. (33) we have a particular solution of the equation as

$$Z(x,t) = Z_0 e^{-ikx} E_{2\alpha}(-\Omega_f^2 t^{2\alpha}) \tag{42}$$

We can easily see that in the case $\alpha = 1$, the solution to the equation is

$$Z(x,t) = \text{Re}(Z_0 e^{i(\Omega t - kx)}) \tag{43}$$

which defines a periodic, with fundamental period $T = 2\pi\Omega$, monochromatic wave in the, $x$, direction and in time, $t$. This result is very well known from the ordinary electromagnetic waves theory. However, for the arbitrary case of $\alpha$ ($0 < \alpha < 1$) the solution is periodic only respect to x and it is not periodic with respect to $t$. The solution represents a plane wave with time decaying amplitude. For example, for the case $\alpha = \frac{1}{2}$ we have

$$u(t) = E_1(-\eta\Omega^2 t) = e^{-\eta\Omega^2 t} \tag{44}$$

Therefore, the solution is

$$Z(x,t) = (Z_0 e^{-\eta\Omega^2 t})e^{-ikx} \tag{45}$$

Also, for the case of $\alpha = \frac{1}{4}$ we have

$$u(t) = E_{\frac{1}{2}}(-\eta^{\frac{3}{2}}\Omega^2 \sqrt{t}) = e^{\eta^3\Omega^4 t}(1+erf(-\eta^{\frac{3}{2}}\Omega^2\sqrt{t})) = e^{\eta^3\Omega^4 t} erfc(\eta^{\frac{3}{2}}\Omega^2\sqrt{t}) \tag{46}$$

where *erfc* denotes the complimentary error function and the error function is defined as

$$erf(z) = \frac{2}{\sqrt{\pi}}\int_0^z e^{-t^2} dt, \quad erfc(z) = 1 - erf(z), \quad z \in C \tag{47}$$

For large values of $z$, the complimentary error function can be approximated as

$$erfc(z) \simeq \frac{1}{\sqrt{\pi z}} \exp(-z^2) \tag{48}$$

Substituting Eq. (46) into Eq. (33) leads to the solution

$$Z(x,t) = (Z_0 e^{\eta^3\Omega^4 t} erfc(\eta^{\frac{3}{2}}\Omega^2\sqrt{t}))e^{-ikx} \tag{49}$$

At asymptotically large times, and using Eq. (39) we have

$$Z(x,t) \simeq (\frac{Z_0}{\sqrt{\pi}\eta^{\frac{3}{2}}\Omega^2\sqrt{t}})e^{-ikx} \tag{50}$$

Then for these cases, the solutions are periodic only respect to $x$ and they are not periodic with respect to $t$. In fact, solutions represent plane waves with time decaying amplitude. Also, for the asymptotic behavior of the solution see the Appendix B.

## 4. Fractional Order Metamaterials

As the first application of the fractional electrodynamics here we briefly introduce the new concept of fractional order metamaterials (FOMs). FOMs are in fact metamaterials [21-25] composed of unit cells which can be modeled using the fractional order electric and electronic elements. Recently we have proposed the notion of fractional electromagnetic metamaterials [2] based on the framework of fractional electrodynamics [1] for describing nonlocal phenomena in electromagnetic metamaterials. Also, in the last decade fractional calculus has found many applications in physics, electrical and electronic engineering. Among these applications here we can mention fractional order modeling of: signals and systems [26] semiconductor diodes [27,28], anomalous charge transport in nanosystems [29], thermal modeling and temperature estimation of a transistor junction [30], electrical circuits and their applications [31- 40] and so on. In this work we want to propose the notion of using fractional order electric and electronic elements in the modeling new type of complex metamaterials which we call them fractional order metamaterials (FOMs). The following figure (Fig. (1)) shows general structure of the FOM which is composed of unit cells of fractional order elements.

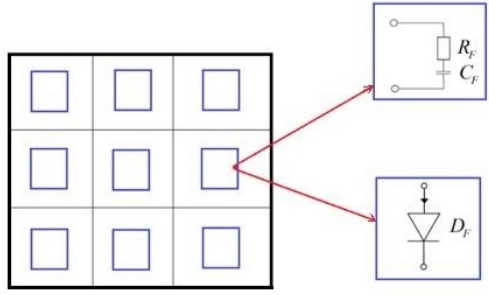

**Figure1:** schematic of FOM constituent elements [3].

As the above figure shows a FOM is composed of complex unit cells which are equivalent to the fractional order elements instead of the conventional electrical and electronic elements. Fractional order resistor, capacitor and inductor are elements whose governing equations are respectively as follows [36]:

$$i(t) = \frac{1}{R_F} D^{1-\alpha} v(t) \tag{51}$$

$$q(t) = C_F D^{1-\beta} v(t) \tag{52}$$

$$v(t) = L_F D^{\gamma} i(t) \tag{53}$$

where in the above equations $D^{\mu}$ is the fractional derivative of order $0 < \mu \leq 1$, $i(t)$ is the total current, $v(t)$, difference voltage of the element, $R_F$, resistance of the fractional resistor, $q(t)$, the total charge of the fractional capacitor and $C_F$ is its capacity and finally $L_F$ is inductance of the fractional order inductor. Since the electromagnetic characteristics of the metamaterials depend on their unit cells components in addition to their shape, size, orientation, and alignment of unit cells, we expect completely different physical characteristics in comparison with the conventional metamaterials and metasurfaces. In our previous work [2] we proposed a theoretical framework to study the electromagnetic properties of such new type of metamaterials which exhibit nonlocal properties and those which cause anomalous light transport and show exotic light-matter interactions [41-43].

The above-mentioned fractional order circuit theory modeling approach can provide an accurate model for the future complex and nonlinear metamaterials working in radio-frequency, microwave or even in higher frequency applications and also for some new nonlinear optical metamaterials. As a special example of this approach, we can use it for the modeling of a complex nonlinear metacrystal formed from resonant circuit elements that couple strongly to the magnetic field by the split ring resonator (SRR) medium [44 and Refs therein] with its equivalent inductively driven RLC circuit model shown in Fig. (2).

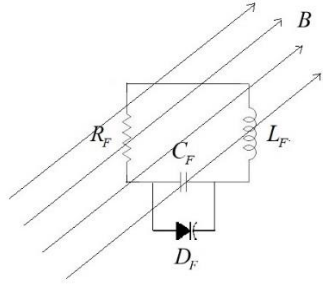

**Figure2:** Equivalent effective fractional circuit model for SRR [3].

The produced current $I_\alpha(t)$ due to the induced electromotive force $\varepsilon_\alpha(t)$, satisfies:

$$L_F D^\alpha I_\alpha + R_F I_\alpha + V_{D_F} = \varepsilon_\alpha(t) = -D^\alpha \Phi_m \tag{54}$$

where $L_F, R_F, V_{D_F}$ are, respectively, the distributed inductance, distributed resistance and the induced voltage across the effective capacitor $C_F$ of the circuit.

### 4-1. Fractional Electromagnetic Metamaterials

In this section first we review different standard model for wave propagation in metamaterials and then we present our new model based on FC. The lossy Drude model [45 and the references therein] is a popular model for metamaterial and in frequency domain can be written in the form of:

$$\varepsilon(\omega) = \varepsilon_0 (1 - \frac{\omega_{pe}^2}{\omega(\omega - i\Gamma_e)}) = \varepsilon_0 \varepsilon_r \tag{55}$$

$$\mu(\omega) = \mu_0 (1 - \frac{\omega_{pm}^2}{\omega(\omega - i\Gamma_m)}) = \mu_0 \mu_r \tag{56}$$

where the parameters $\varepsilon_0$ and $\mu_0$ are the vacuum electric permittivity and magnetic permeability, the parameters $\omega_{pe}$ and $\omega_{pm}$ are the electric and magnetic plasma frequencies, $\Gamma_e$ and $\Gamma_m$ are the electric and magnetic damping frequencies, and $\omega$ is a general frequency, respectively. Based on the above model for metamatrials we can obtain the governing equations for modeling the wave propagation in metamatrials described by the Drude model as follow:

$$\varepsilon_0 \partial_t \vec{E} + \vec{J} = \vec{\nabla} \times \vec{H} \tag{57}$$

$$\mu_0 \partial_t \vec{H} + \vec{K} = -\vec{\nabla} \times \vec{E} \tag{58}$$

$$\frac{1}{\varepsilon_0 \omega_{pe}^2} \partial_t \vec{J} + \frac{\Gamma_e}{\varepsilon_0 \omega_{pe}^2} \vec{J} = \vec{E} \tag{59}$$

$$\frac{1}{\mu_0 \omega_{pm}^2} \partial_t \vec{K} + \frac{\Gamma_m}{\mu_0 \omega_{pm}^2} \vec{K} = \vec{H} \tag{60}$$

where $\vec{E}$ and $\vec{H}$ are the electric and magnetic fields, respectively, $\vec{J}$ and $\vec{K}$ are the polarization and magnetization current densities, respectively.

And for the so-called Lorentz model for modeling wave propagation in metamaterials, we have [45 and the references therein]:

$$\varepsilon(\omega) = \varepsilon_0 (1 - \frac{\omega_{pe}^2}{\omega^2 - \omega_{e0}^2 - i\Gamma_e \omega}) \tag{61}$$

$$\mu(\omega) = \mu_0 (1 - \frac{\omega_{pe}^2}{\omega^2 - \omega_{m0}^2 - i\Gamma_m \omega}) \tag{62}$$

where $\omega_{e0}$ and $\omega_{m0}$ are the electric and magnetic resonance frequencies, respectively.

The above model will result in the following Lorentz model equations for metamaterials:

$$\varepsilon_0 \partial_t \vec{E} + \vec{J} = \vec{\nabla} \times \vec{H} \tag{63}$$

$$\mu_0 \partial_t \vec{H} + \vec{K} = -\vec{\nabla} \times \vec{E} \tag{64}$$

$$\frac{1}{\varepsilon_0 \omega_{pe}^2} \partial_t \vec{J} + \frac{\Gamma_e}{\varepsilon_0 \omega_{pe}^2} \vec{J} + \frac{\omega_{e0}^2}{\varepsilon_0 \omega_{pe}^2} \vec{P} = \vec{E} \tag{65}$$

$$\frac{1}{\mu_0 \omega_{pm}^2} \partial_t \vec{K} + \frac{\Gamma_m}{\mu_0 \omega_{pm}^2} \vec{K} + \frac{\omega_{m0}^2}{\mu_0 \omega_{pm}^2} \vec{M} = \vec{H} \tag{66}$$

where $\vec{P}$ and $\vec{M}$ are the polarization and magnetization vectors, respectively and polarization and magnetization current densities can be defined using them as:

$$\vec{J} = \frac{\partial \vec{P}}{\partial t} \quad , \quad \vec{K} = \frac{\partial \vec{M}}{\partial t} \tag{67}$$

And finally, we have the Drude-Lorentz model in which the permittivity is described by the Drude model, while the permeability is described by the Lorentz model i.e., we have [45 and the references therein]:

$$\varepsilon(\omega) = \varepsilon_0 (1 - \frac{\omega_p^2}{\omega(\omega + iv)}) \tag{68}$$

$$\mu(\omega) = \mu_0 (1 - \frac{F \omega_0^2}{\omega^2 + i\gamma\omega - \omega_0^2}) \tag{69}$$

where $\omega$ is the excitation angular frequency, $\omega_p$ is the effective plasma frequency, and $v \geq 0$ is the loss parameter, $\omega_0$ is the resonant frequency, $\gamma \geq 0$ is the loss parameter, and $F \in (0,1)$ is a parameter depending on the geometry of the unit cell of the metamaterial. Based on the above model we can write the governing equations for the Drude-Lorentz model as following:

$$\varepsilon_0 \partial_t \vec{E} + \vec{J} = \vec{\nabla} \times \vec{H} \tag{70}$$

$$\mu_0 \partial_t \vec{H} + \vec{K} = -\vec{\nabla} \times \vec{E} \tag{71}$$

$$\frac{1}{\mu_0 \omega_0^2 F} \partial_t \vec{K} + \frac{\gamma}{\mu_0 \omega_0^2 F} \vec{K} + \frac{1}{\mu_0 F} \vec{M} = \vec{H} \tag{72}$$

$$\frac{1}{\varepsilon_0 \omega_p^2} \partial_t \vec{J} + \frac{v}{\varepsilon_0 \omega_p^2} \vec{J} = \vec{E} \tag{73}$$

In the following using the fractional vector differential vectors' definitions presented in previous section and the above governing equations for modeling the wave propagation in metamatrials described by the Drude model, we introduce new fractional Drude metamaterial model which contains intrinsically nonlocal operators instead of local ones as a model for description of electromagnetic properties of recent nonlocal metamaterials. For this purpose, we consider two different cases:

**4-1-1. Case a:** $0 < \alpha, \beta < 1$

For the first case we consider Maxwell's equations in a Drude metamaterial [45, 46] in the framework of fractional calculus as:

$$\varepsilon_0 \partial_t^\alpha \vec{E} + \vec{J} = (\vec{\nabla} \times)^\beta \vec{H} \tag{74}$$

$$\mu_0 \partial_t^\alpha \vec{H} + \vec{K} = -(\vec{\nabla} \times)^\beta \vec{E} \tag{75}$$

$$\partial_t^\alpha \vec{J} = \varepsilon_0 \omega_{pe}^{2\alpha} \vec{E} \tag{76}$$

$$\partial_t^\alpha \vec{K} = \mu_0 \omega_{pm}^{2\alpha} \vec{H} \tag{77}$$

In the case of low level fractionality [47, 48] i.e., when the order of fractional calculus is close to integer i.e., the case of $\alpha = n - \varepsilon$ where $\varepsilon \ll 1$ we have:

**Case 1:** Caputo fractional derivative of order $\alpha = 1 - \varepsilon$
For this case we have:
$${}^C\partial_t^{1-\varepsilon} f(t) = \partial_t f + \varepsilon(\partial_t f(0)\ln(t) + ...) \tag{78}$$

**Case 2:** Caputo fractional derivative of order $\alpha = 2 - \varepsilon$
For this case we have:
$${}^C\partial_t^{2-\varepsilon} f(t) = (\partial_t)^2 f + \varepsilon((\partial_t)^2 f(0)\ln(t) + ...) \tag{79}$$

**Case 3:** Riesz fractional derivative of order $\alpha = 2 - \varepsilon$
For this case we have:
$${}^R\partial_{|x|}^{2-\varepsilon} f(x) = (\partial_{|x|})^2 f + \varepsilon(\gamma(\partial_{|x|})^2 f + ...) \tag{80}$$

where $\gamma = 0.577215664901532...$ Euler–Mascheroni constant.

Base on the above equations one can find that the fractional Drude metamaterial model is characterized by a dispersive permittivity and permeability given as:

$$\varepsilon(\omega) = \varepsilon_0 \left(1 - \frac{\omega_{pe}^{2\alpha}}{\omega^{2\alpha}}\right), \quad \mu(\omega) = \mu_0 \left(1 - \frac{\omega_{pm}^{2\alpha}}{\omega^{2\alpha}}\right) \tag{81}$$

We call systems of Eqs. (74-77) the fractional Drude metamaterial model of order $\alpha$ and $\beta$ in which $0 < \alpha, \beta < 1$.

**4-1-2. Case b:** $1 < \alpha < 2$, $0 < \beta < 1$

System of Eqs. (74-77) can be rewritten in the higher fractional orders i.e., $1 < \alpha < 2$, $0 < \beta < 1$ and with the condition of low level fractionality Eqs. (78, 79), in which we get two sets of equations:

- the first pair of equations involving only the field variables ($\vec{E}$, $\vec{K}$) modeled by the system of fractional equations:

$$\partial_t^\alpha \vec{E} + c^2 (\vec{\nabla} \times)^\beta ((\vec{\nabla} \times)^\beta \vec{E}) + \omega_{pe}^2 \vec{E} = -c^2 (\vec{\nabla} \times)^\beta \vec{K} \tag{82}$$

$$\partial_t^\alpha \vec{K} + \omega_{pm}^{2\alpha} \vec{K} = -\omega_{pm}^{2\alpha} (\vec{\nabla} \times)^\beta \vec{E} \tag{83}$$

- the second ones involving the field variables ($\vec{H}$, $\vec{J}$) modeled by the system of fractional equations:

$$\partial_t^\alpha \vec{H} + c^2 (\vec{\nabla} \times)^\beta ((\vec{\nabla} \times)^\beta \vec{E}) + \omega_{pm}^{2\alpha} \vec{H} = c^2 (\vec{\nabla} \times)^\beta \vec{J} \tag{84}$$

$$\partial_t^\alpha \vec{J} + \omega_{pe}^{2\alpha} \vec{J} = \omega_{pe}^{2\alpha} (\vec{\nabla} \times)^\beta \vec{H} \tag{85}$$

We note that we derived these systems of decupled equations through the divergence conditions.

## 5. Fotonics

Fotonics deals with optical and photonic phenomena through natural and engineered artificial complex media in the framework of fractional dynamics. As an example of its application, we investigate the physics of random lasers in this framework.

### 5-1. Fractional order random laser

In conventional lasers, the optical cavity determines essential characteristics of the lasing modes however in spite this fact that random lasers work on the same principles, but the modes are determined by multiple scattering and not by a laser cavity [49-54]. In recent years the powerful framework of fractional dynamics has found many applications in science and engineering and in particular in description of wave propagation in complex media with disorder [55-58]. More recently the idea of fractional order random laser (FORL) has been proposed in [59] in which the authors used the classical Letokhov model of a non-resonant random laser that is formulated in terms of the reaction-diffusion equation obeyed by the optical energy density $W(\vec{r}, t)$ as:

$$\frac{\partial W(\vec{r}, t)}{\partial t} = D \nabla^2 W(\vec{r}, t) + \frac{v}{l_g} W(\vec{r}, t) \tag{86}$$

where $D$ is the diffusion constant of photons given by $D = \frac{vl_t}{2n}$, where $l_t$ is the transport mean free path and $n$ is the dimensionality of the problem, $v$ is the speed of light in the medium, and $l_g$ ($l_g \gg l_t$) is its characteristic gain length and then based on time fractional generalization of Letokhov diffusion model as:

$$^C D_{\tilde{t}}^\alpha \tilde{W}(x,\tilde{t}) = K_\alpha \tau_d^\alpha \frac{\partial^2 \tilde{W}(x,\tilde{t})}{\partial x^2} + (\frac{\tau_d}{\tau_g})^\alpha \tilde{W}(x,\tilde{t}) \tag{87}$$

wherein we will consider the scaling $\tau_d = \frac{l_t}{v}$ and $\tau_g = \frac{l_g}{v}$, which are the characteristic time for the scattering and the amplification time of a photon, respectively, $l_t$ is the transport mean free path, the scaled time variable is $\tilde{t} = \frac{t}{\tau_d}$, $K_\alpha = \frac{\Gamma(\alpha+1)\langle x^2(t)\rangle}{2t^\alpha}$ ($\langle x^2 \rangle$ is the average square displacement)

So, we can easily find the time dependent part of the solution of Eq. (87), $f(t)$, amplification length, $L_\alpha$ and consequently the lasing volume $V_\alpha$ as:

$$f(t) = E_\alpha([(\frac{\tau_d}{\tau_g})^\alpha - (\frac{n\pi}{L_\alpha})^2 K_\alpha \tau_d^\alpha]t^\alpha), \quad L_\alpha = n\pi\sqrt{K_\alpha(\frac{l_g}{v})^\alpha}, \quad V_\alpha = n^3\pi^3 K_\alpha^{\frac{3}{2}}(\frac{l_g}{v})^{\frac{3\alpha}{2}} \tag{88}$$

They are also proposed the superdiffusive regime of the FORL by considering the reaction-diffusion space-fractional equation as [59]:

$$\frac{\partial W(x,t)}{\partial t} = K_\beta\, ^R D_{|x|}^\beta W(x,t) + \frac{v}{l_g} W(x,t) \tag{89}$$

where $^R D_{|x|}^\beta$ is the Riesz fractional derivatives of order $\beta$ ($1 < \beta < 2$) defined as:

(90)

$$^R D_{|x|}^\beta g(x) = -\frac{1}{2\cos(\frac{\pi\beta}{2})\Gamma(m-\beta)} \cdot \frac{d^m}{dx^m}\left(\int_{-\infty}^{t}(x-\chi)^{m-\beta-1}g(\chi)d\chi + (-1)^m\int_{t}^{\infty}(x-\chi)^{m-\beta-1}g(\chi)d\chi\right)$$

where $m-1 \leq \beta < m \in Z^+$, which in the case of $\beta = 2-\varepsilon$ is equal to [14]:

$$^R D_{|x|}^{2-\varepsilon} g(x) = (D_{|x|})^2 g + \varepsilon(\gamma(D_{|x|})^2 g + ...). \tag{91}$$

The above-mentioned approach for the subdiffusive and superdiffusive optical gain media can be considered as a theoretical framework for the FORL. Also, a theoretical model for random lasers based on the superdiffusive optical gain medium has been presented in [60] in which the authors considered the generalized Lambert-Beer law as:

$$p(l) = \sigma^{-\beta}l^{(\beta-1)}E_{\beta,\beta}(-(\frac{l}{\sigma})^\beta) \tag{92}$$

where $\sigma$ is a scattering characteristic length that decreases as the scattering of the medium becomes stronger, $l$ is the step lengths and $E_{\beta,\beta}$ is the two-parameter Mittag-Leffler function.

We can here generalize the above theoretical models in several possible ways:

**Case1**: space-time fractional Letokhov model for the FORL:

$$^{C}D_{\tilde{t}}^{\alpha}\tilde{W}(x,\tilde{t}) = \left(K_{\alpha,\beta}\tau_{d}^{\alpha}\right){}^{R}D_{|x|}^{\beta}\tilde{W}(x,\tilde{t}) + (\frac{\tau_{d}}{\tau_{g}})^{\alpha}\tilde{W}(x,\tilde{t}) \qquad (93)$$

**Case2:** variable order space-time fractional Letokhov model for the FORL: because of the interaction between emitted light and the medium and regarding to this fact that the order of fractional operators can be considered as an index of the complexity of the medium we can consider the cases of $\alpha = \alpha(t)$ and $\beta = \beta(x,t)$ which in these cases fractional order derivative reads as [61]:

$$_{0}^{c}D_{t}^{\alpha(t)}f(t) = \frac{d^{n}}{dt^{n}}\left(\frac{1}{\Gamma(n-\alpha(t))}\int_{0}^{t}(t-\tau)^{n-\alpha(t)-1}f(\tau)d\tau\right) \qquad (94)$$

with the following property:

$$_{a}^{c}D_{t}^{\alpha(t)}(t-a)^{\gamma} = \frac{\Gamma(\gamma+1)}{\Gamma(\gamma-\alpha(t)+1)}(t-a)^{\gamma-\alpha(t)} \qquad (95)$$

$$-\alpha'(t)\frac{\Gamma(\gamma+1)}{\Gamma(\gamma-\alpha(t)+2)}(t-a)^{\gamma-\alpha(t)+1}\times[\ln(t-a)-\Psi(\gamma-\alpha(t)+2)]$$

where $\Psi$ is the Psi function which is equal to the derivative of the logarithm of the Gamma function i.e.:

$$\Psi(t) = \frac{d}{dt}\ln(\Gamma(t)) = \frac{\Gamma'(t)}{\Gamma(t)} \qquad (96)$$

**Case3:** random lasing can be produced using some new fractal media such as fractal perovskite thin films [62]. So, we can propose a new model based on the Hausdorff derivative which successfully can describe processes in complex fractal and porous media [63-65] and is defined as:

$$^{H}D_{t}^{\alpha}f(t) = \lim_{t\to t'}\frac{f(t)-f(t')}{t^{\alpha}-t'^{\alpha}} = \frac{1}{\alpha t^{\alpha-1}}\frac{df}{dt} \qquad (97)$$

where $t$ and $t'$, respectively, represent the final and internal time instances and $\alpha$ is the time fractal dimensionality. So, a fractal Letokhov model can be considered as a promising theoretical framework for the random lasing due to such fractal structures and future fractal random laser as:

$$^{H}D_{t}^{\alpha}\tilde{W}(x,\tilde{t}) = \left(K_{\alpha,\beta}\tau_{d}^{\alpha}\right){}^{H}D_{x}^{\beta}\tilde{W}(x,\tilde{t}) + (\frac{\tau_{d}}{\tau_{g}})^{\alpha}\tilde{W}(x,\tilde{t}) \qquad (98)$$

## 6. Conclusion

Fractional electromagnetic field theory describes electromagnetic wave propagation through the complex, nonlocal, dissipative, fractal and also recent artificially engineered materials know as fractional metamaterials. In this theory using the fractional Maxwell's equations we are also able to consider the well-known effect of electromagnetic memory. In this review we investigated some applications of the powerful theory of fractional electrodynamics including: fractional order metamaterials and fractional order random laser. Also, some modern approaches to the fractional electromagnetic field theory are still in progress, see the Appendix C for the gauge invariant fractional electrodynamics.

## Appendix A: Fractional dynamics of charged particles

For the simplest case we can consider motion of charged particles in a uniform electric field $\vec{E} = E_z \hat{k}$. So, using the fractional Newton's second law we have

$$^c_0 D_t^\alpha p_z(t) = \eta^{1-\alpha} q E_z \tag{99}$$

if $p_z(0) = 0$, so we have

$$p_z(t) = \frac{q E_z \eta^{1-\alpha}}{\Gamma(\alpha+1)} t^\alpha \tag{100}$$

where $p_z$ is the z-component of particle's momentum. Also, we can easily calculate z-component of particle's position as a function of time, i.e., $z(t)$ from

$$p_z(t) = \frac{m}{\eta^{1-\alpha}} (^c_0 D_t^\alpha z(t)) \tag{101}$$

Taking into account the initial condition $z(0) = 0$ and substituting Eq. (100) into Eq. (101) leads to the solution

$$z(t) = \frac{q E_z \eta^{2-2\alpha}}{m \Gamma(2\alpha+1)} t^{2\alpha} \tag{102}$$

For the case of $\alpha = 1$ we can easily show that

$$z_{\alpha=1}(t) = \frac{q E_z}{m \Gamma(3)} t^2 = \frac{q E_z}{2m} t^2 \tag{103}$$

as expected from the standard electrodynamics.

## Appendix B: Asymptotic behavior of the solution

The algebraic decay of the solutions of the fractional equations is the most important effect of the fractional derivative in the typical fractional equations contrary to the exponential decay of the usual standard form of the equations. To describe this algebraic decay in our case, we consider the integral form for the Mittag-Leffler function. The asymptotic expansion of $E_\alpha(z)$ based on the integral representation of the Mittag-Leffler function in the form [66]

$$E_\alpha(z) = \frac{1}{2\pi i} \int_\Upsilon \frac{t^{\alpha-1} \exp(t)}{t^\alpha - z} dt \tag{104}$$

where $\Re(\alpha) > 0$, $(\alpha, z) \in C$ and the path of integration $\Upsilon$ is a loop starting and ending at $-\infty$ and encircling the circular disk $|t| \leq |z|^{\frac{1}{\alpha}}$ in the positive sense $|\arg t| < \pi$ on $\Upsilon$. The integrand has a branch point at $t=0$. The complex $t$-plane is cut along the negative real axis and in the cut plane the integrand is single-valued the principal branch of $t^\alpha$ is taken in the cut plane. Eq. (104) can be proved by expanding the integrand in powers of $t$ and integrating term by term by making use of the well-known Hankel's integral for the reciprocal of the gamma function, namely

$$\frac{1}{\Gamma(\beta)} = \frac{1}{2\pi i} \int_{H_a} \frac{e^\zeta}{\zeta^\beta} d\zeta \tag{105}$$

The integral representation Eq. (104) can be used to obtain the asymptotic expansion of the Mittag-Leffler function at infinity. Accordingly, the following cases are obtained.

If $0 < \alpha < 2$ and $\mu$ is a real number such that $\frac{\pi\alpha}{2} < \mu < \min[\pi, \pi\alpha]$ then for $N^* \in N$, $N^* \neq 1$ there holds the following asymptotic expansion:

$$E_\alpha(z) = \frac{1}{\alpha} z^{\frac{(1-\beta)}{\alpha}} \exp(z^{\frac{1}{\alpha}}) - \sum_{r=1}^{N^*} \frac{1}{\Gamma(1-\alpha r)} \frac{1}{z^r} + O[\frac{1}{z^{N^*+1}}] \tag{106}$$

as $|z| \to \infty$, $|\arg z| \leq \mu$ and

$$E_\alpha(z) = -\sum_{r=1}^{N^*} \frac{1}{\Gamma(1-\alpha r)} \frac{1}{z^r} + O[\frac{1}{z^{N^*+1}}] \tag{107}$$

as $|z| \to \infty$, $\mu \leq |\arg z| \leq \pi$. In our case, $z = -\Omega_f^2 t^{2\alpha}$ and

$$E_{2\alpha}(-\Omega_f^2 t^{2\alpha}) \simeq \frac{1}{\Gamma(1-2\alpha)} \frac{1}{\Omega_f^2 t^{2\alpha}} \tag{108}$$

Then, substitution of Eq. (108) into Eq. (42) gives

$$Z(x,t) \simeq [(\frac{Z_0}{\Omega_f^2 \Gamma(1-2\alpha)}) t^{-2\alpha}] e^{-ikx} \tag{109}$$

As we can see in this result, we arrive to the asymptotic solution for the electromagnetic wave equation which represents a plane wave with algebraic time-decaying amplitude. This is a direct consequence of the fractional time derivative in the system. In the other word fractional differentiation with respect to time can be interpreted as an existence of memory effects which correspond to intrinsic dissipation in our system.

**Appendix C: Gauge invariant fractional electrodynamics**
Recently gauge invariant fractional electromagnetic field has been proposed [67]. In this work the author have used well-known left (forward) and right (backward) Riemann–Liouville (RL) partial fractional derivatives (which respectively are nonlocal causal operators and nonlocal non-causal operators) of order $\alpha_\mu$, $\beta_\mu$ (which are positive real or even complex numbers) of a real valued function $f$ of $d+1$ real variables $x^0, x^1, ..., x^d$ with respect to $x_\mu$, i.e.:

$$_{a_\mu}\partial_\mu^{\alpha_\mu} f(x^0,...,x^d) = \frac{1}{\Gamma(n_\mu - \alpha_\mu)} \partial_{x^\mu}^{n_\mu} \int_{a_\mu}^{x^\mu} \frac{f(x^0,...,x^{\mu-1},u,x^{\mu+1},...,x^d)}{(x^\mu - u)^{1+\alpha_\mu - n_\mu}} du \quad \text{(left RL)} \tag{110}$$

$$_\mu\partial_{b_\mu}^{\beta_\mu} f(x^0,...,x^d) = \frac{(-1)^{n_\mu}}{\Gamma(n_\mu - \beta_\mu)} \partial_{x^\mu}^{n_\mu} \int_{x}^{b_\mu} \frac{f(x^0,...,x^{\mu-1},u,x^{\mu+1},...,x^d)}{(u - x^\mu)^{1+\beta_\mu - n_\mu}} du \quad \text{(right RL)} \tag{111}$$

where $\partial_{x^\mu}^{n_\mu}$ is the ordinary partial derivative of integer order n with respect to the variable $x$ and $a_\mu$, $b_\mu$ are real number which define the domain. It is worth mentioning that only left (forward) RL operator will be resulted in a causal theory and right (backward) RL operator will produce a non-causal theory. Based on above definitions one can introduce the left–right (two-sided backward-forward) fractional Riemann–Liouville operators i.e.:

$$\partial_\mu^{\alpha\beta} = \frac{1}{2}({}_{a_\mu}\partial_\mu^{\alpha_\mu} - {}_\mu\partial_{b_\mu}^{\beta_\mu}) \tag{112}$$

where for the case of $\alpha_\mu = \beta_\mu$ becomes the well-known Riesz fractional derivatives which its explicit form for the special case of $1 < \alpha_i < 2$ becomes:

$$\partial_\mu^{\alpha\alpha} f(x^0,...,x^d) = \frac{-1}{2\cos(\frac{\pi\alpha_\mu}{2})\Gamma(2-\alpha_\mu)} \partial_{x^\mu}^2 \int_{a_\mu}^{b_\mu} \frac{f(x^0,...,x^{\mu-1},u,x^{\mu+1},...,x^d)}{|x^\mu - \xi|^{\alpha_\mu - 1}} d\xi \tag{113}$$

By use of the above relations, we will briefly review the recently proposed gauge invariant fractional electrodynamics [1,67,68]. For this purpose, we start from the lagrangian in the form of $L = L(A_\mu, \partial_\mu^{\alpha\beta} A_\mu, x^\mu)$ for $N$ fields $A_\mu \equiv A_\mu(x^0,...,x^d)$ where $\mu = 1, 2, ..., N$. So, for $0 < \alpha_\mu, \beta_\mu < 1$ we will have the following fractional Euler–Lagrange equation (see Appendix D for details) for $A_\mu$ fields:

$$\frac{\partial L}{\partial A_\mu} - \partial_\nu^{\beta\alpha} \frac{\partial L}{\partial(\partial_\nu^{\alpha\beta} A_\mu)} = 0 \tag{114}$$

Now suppose $A_\mu = (\varphi, -\vec{A})$ as the 4-vector electromagnetic potential, then immediately we can write the standard Maxwell's field strength tensor $F_{\mu\nu}$ as: $F_{\mu\nu} = \partial_\mu A_\nu - \partial_\nu A_\mu$ and so its direct fractional generalization will take the form of: $F_{\mu\nu}^\alpha = {}_\mu\partial_b^\alpha A_\nu - {}_\nu\partial_b^\alpha A_\mu$ which has been written in terms of non-causal right Riemann–Liouville derivatives and with the following matrix form of:

$$F_{\mu\nu}^\alpha = \begin{pmatrix} 0 & E_x^\alpha & E_y^\alpha & E_z^\alpha \\ -E_x^\alpha & 0 & -B_z^\alpha & B_y^\alpha \\ -E_y^\alpha & B_z^\alpha & 0 & -B_x^\alpha \\ -E_z^\alpha & -B_y^\alpha & B_x^\alpha & 0 \end{pmatrix} \tag{115}$$

and its contravariant version in the form of:

$$F_\alpha^{\mu\nu} = \begin{pmatrix} 0 & E_\alpha^x & E_\alpha^y & E_\alpha^z \\ -E_\alpha^x & 0 & -B_\alpha^z & B_\alpha^y \\ -E_\alpha^y & B_\alpha^z & 0 & -B_\alpha^x \\ -E_\alpha^z & -B_\alpha^y & B_\alpha^x & 0 \end{pmatrix} \qquad (116)$$

where $F_{0i}^\alpha$ and $F_{ij}^\alpha$ components are defined as the fractional electric fields fractional magnetic fields respectively (Latin and Greek indices (excluding $\alpha, \beta$) are respectively equal to: $i, j, k = 1, 2, 3$ and $\mu, \nu = 0, 1, 2, 3$). Finally, we can write the Lagrangian density of the fractional electromagnetic field in terms of the fractional Maxwell's field strength tensor $F_{\mu\nu}^\alpha$ as:

$$L_{\text{Fractional Electromagnetic Field}} = -\frac{1}{16\pi c} F_{\mu\nu}^\alpha F_\alpha^{\mu\nu} \qquad (117)$$

where $F_\alpha^{\mu\nu} = \eta^{\mu\rho} \eta^{\nu\sigma} F_{\rho\sigma}^\alpha$ and the Minkowski metric as: $\eta^{\mu\nu} = \eta_{\mu\nu} = diag(+1, -1, -1, -1)$.

Using the above lagrangian and the fractional Euler–Lagrange equations and the condition of $\partial_{[\rho}^\alpha F_{\mu\nu]}^\alpha = 0$ which $[\rho\mu\nu]$ denotes the antisymmetrized sum over permutations of the indices $\rho$, $\mu$, and $\nu$ for the fractional Maxwell's field strength tensor we can easily derive the other form of Maxwell's equations in terms of only causal left fractional derivatives as:

$$_a\partial_\nu^\alpha F_\alpha^{\mu\nu} = -\frac{4\pi}{c} j^\mu \qquad (118)$$

**Appendix D: Fractional Euler-Lagrange Equation**

The fractional variational principle represents an important part of fractional calculus and has found many applications in physics. As it is mentioned in [69-71] there are several versions of fractional variational principles and fractional Euler-Lagrange equations due to the fact that we have several definitions for the fractional derivatives. In this work we use new approach presented in [70] where authors developed the action principle for field systems described in terms of fractional derivatives, by use of a functional $S(\phi)$ as:

$$S(\phi) = \int L\left(\phi(x_k), \left(_+^C\partial_k^\alpha\right)\phi(x_k), \left(_-^C\partial_k^\alpha\right)\phi(x_k), x_k\right)(dx_k) \qquad (119)$$

where $L\left(\phi(x_k), \left(_+^C\partial_k^\alpha\right)\phi(x_k), \left(_-^C\partial_k^\alpha\right)\phi(x_k), x_k\right)$ is a Lagrangian density function. Accordingly, $x_k$ represents $n$ variables $x_1$ to $x_n$, $\phi(x_k) \equiv \phi(x_1, ..., x_n)$, $L\left(*, _+^C\partial_k^\alpha, *, *\right) \equiv L\left(*, _+^C\partial_1^\alpha, ..., _+^C\partial_n^\alpha, *, *\right)$, $(dx_k) \equiv dx_1...dx_n$ and the integration is taken over the entire domain $\Omega$. From these definitions, we can obtain the fractional Euler-Lagrange equation as:

$$\frac{\partial L}{\partial \phi} + \sum_{k=1}^n {}_-\partial_k^\alpha \frac{\partial L}{\partial \left(_+^C\partial_k^\alpha \phi\right)} + \sum_{k=1}^n {}_+\partial_k^\beta \frac{\partial L}{\partial \left(_-^C\partial_k^\beta \phi\right)} = 0 \qquad (120)$$

Above equation is the Euler–Lagrange equation for the fractional field system and for $\alpha, \beta \to 1$, gives the usual Euler–Lagrange equations for classical fields. Also, we can study the

Hamiltonian formulation of the field systems [70]. For this, consider the fractional Lagrangian given in equation (119). Then the fractional canonical momentum densities $\pi_{\alpha_k}$ and $\pi_{\beta_k}$ are:

$$\pi_{\alpha_k} = \frac{\partial L}{\partial \left( {}^C_+\partial_k^\alpha \phi \right)}, \quad \pi_{\beta_k} = \frac{\partial L}{\partial \left( {}^C_-\partial_k^\beta \phi \right)} \tag{121}$$

So the fractional canonical Hamiltonian density is:

$$H = \sum_{k=1}^{n} \left( \pi_{\alpha_k} \left( {}^C_+\partial_k^\alpha \phi \right) + \pi_{\beta_k} \left( {}^C_-\partial_k^\beta \phi \right) \right) - L \tag{122}$$

With this fractional canonical Hamiltonian density one can easily obtain the fractional Hamilton equation of motion as:

$$\frac{\partial H}{\partial \phi} = \sum_{k=1}^{n} \left( {}_-\partial_k^\alpha \pi_{\alpha_k} + {}_+\partial_k^\beta \pi_{\beta_k} \right) \tag{123}$$

One can prove that above equations can lead to the correct Euler–Lagrange equation of motion.